\documentstyle[12pt]{article}
  \advance\oddsidemargin  by -1in
  \advance\evensidemargin by -1in
\def\lsim{\mathrel{\rlap{\lower4pt\hbox{\hskip1pt$\sim$}}
    \raise1pt\hbox{$<$}}}         
\def\gsim{\mathrel{\rlap{\lower4pt\hbox{\hskip1pt$\sim$}}
    \raise1pt\hbox{$>$}}}         

\def\beq{\begin{equation}}
\def\endeq{\end{equation}}
\def\arr{\begin{eqnarray}}
\def\endarr{\end{eqnarray}}
\def\bm{\boldmath}

\begin{document}

\pagestyle{empty}

\begin{flushright}
DFTT 72/93\\
hep-ph/9403236\\
January 1994
\end{flushright}

\vspace{2 cm}

\begin{center}
{\bf \Large Leptoproduction of charm revisited}
\vspace{1cm}\\
V.~BARONE$^{a}$, M.~GENOVESE$^{a}$,
N.~NIKOLAEV$^{b,c}$,\\
E.~PREDAZZI$^{a}$ and B.~ZAKHAROV$^{c}$\\

\vspace{1 cm}

{\sl $^{a}$Dipartimento di Fisica Teorica,
Universit\`a di Torino\\ and INFN, Sezione di Torino, \\
Via P.Giuria 1, I--10125 Torino, Italy\smallskip\\
$^{b}$Institut f{\"u}r Kernphysik, Forschungszentrum J\"ulich,\\ D-52425
J\"ulich, Germany\smallskip\\
$^{c}$L.~D.~Landau Institute for
Theoretical Physics, GSP-1, 117940,\\
ul. Kosygina 2, Moscow V-334, Russia
\vspace{1.2cm}\\ }
\underline{ABSTRACT}\medskip\\
\end{center}

We calculate the energy--momentum
distribution of the charmed quarks produced
in neutrino reactions on protons, quantifying the importance of mass
and current non--conservation effects.
We study the strange and charm distributions probed in
neutrino interactions in the
presently accessible kinematical region.
Some ambiguities inherent to the extraction of the parton densities
from dimuon data are pointed out.

\pagebreak
\pagestyle{plain}


Charm excitation in neutrino scattering is a primary source
of information on the strangeness in nucleons. At moderate values
of $Q^2$ (the momentum transfer squared in deep inelastic scattering),
subtle threshold and current non--conservation effects are
at work \cite{BGNPZ1a,BGNPZ1b,BGNPZ1c,BGNPZ1d},
which lead to the non--universality of the
charm contribution to the structure functions probed by
muons and by neutrinos.
The driving term of the neutrino-excitation
of charm is the $W$-gluon fusion process $W^{+}g \rightarrow
c\bar{s}$ (Fig.~1), which consists of two partonic subprocesses:
excitation of charm on the strange sea and excitation of
anti-strangeness on the anti-charmed sea (we shall
consider only the Cabibbo unsuppressed reactions). Both subprocesses are
of course an integral part of the gauge-invariant $QCD$ cross section
and gauge invariance requires both amplitudes (Figs.~1a,1b)
to be taken into account. In principle, the
two subprocesses can be separated according to which parton,
$c$ or $\bar{s}$, is produced in the $W$-boson
emisphere in the Breit reference frame (other conventions are
possible, as well).
Such
a separation becomes well defined at $Q^{2} \gg (\mu +m)^{2}$,
which is not yet the case with the present experiments (hereafter,
$m$ and $\mu$ will denote the mass of the strange and of the charmed
quark, respectively).

{}From an experimental point of view,
the extraction of what is called the strange structure function of
the nucleon is a multi-stage process (see \cite{CCFR}
and the references therein).
 The signature of the
excitation of the charmed quark is the semileptonic decay
$c\rightarrow s\mu^{+}\nu$, which leads to the opposite-sign muon
pairs $\mu^{+}\mu^{-}$ in the final state (here $\mu^{-}$ comes from
the primary weak interaction vertex). The main source
of background are
the semileptonic decays of pions and kaons. This
forces one to put stringent
lower cuts on the energy of $\mu^{+}$.
Because of these cuts the
measured cross section is sensitive to the energy of the produced
charmed quarks.
In the limit where the masses of the quarks can be neglected,
the primary
produced quark carries $\approx 100\%$ of the $W's$ energy, and
then fragments with some fragmentation function (usually supposed
to be independent of $Q^{2}$)
into the charmed particles, the semileptonic decays
of which produce the desired signature of the charm production.
At moderate values of $Q^{2}$, the mass effects
deeply modify this scenario.
In this paper we shall show that,
at moderate $Q^{2}$,  the primary charmed quarks
have quite a broad energy distribution. Furthermore, this energy
distribution depends on $Q^{2}$, and understanding this dependence
is important for unfolding the acceptance effects and
for the determination
of the primary charm production rate.


We shall focus on the region
of small values of the Bjorken variable $x=Q^{2}/2m_{p}\nu$ and
on the lowest-order QCD subprocesses, which simplifies the analysis.
However, it will be evident that
our main conclusions have a broader applicability.

We have to calculate the $W$- and $\gamma^{*}-$
absorption cross section $d\sigma_{T,L}/dz d^{2}\mbox{\bm $k$}$
for the  transverse ($T$) and
longitudinal ($L$) $W$ and $\gamma^{*}$.
We denote by
$\mbox{\bm $k$}$
the transverse momentum of the charmed quark, and by
\beq
z=\frac{\omega+\sqrt{\omega^{2}-\mu^{2}-
\mbox{\bm $k$}^{2}}}{\nu
+\sqrt{\nu^{2}+Q^{2}}}
\endeq
 the  fraction of the light--cone momentum
of the $W$ $(\gamma^{*})$ carried by the charmed quark. Here
we concentrate on the $(z,\mbox{\bm $k$})$ distribution of the produced
quarks, {\it i.e.} of the jets generated by these quarks,
leaving aside the issue of quark fragmentation into charmed
hadrons.

The analysis of the energy and momentum distribution in the case
of massless quarks and/or asymptotically
large $Q^{2}$ was first performed by Altarelli and Martinelli \cite{AM}.
More recently Ellis and Nason \cite{EN} gave a very detailed treatment
of the real photoproduction of charm. One important finding was
that the higher--order QCD corrections only slightly change
the energy and transverse momentum distribution of the produced quarks.

In this paper we shall be concerned with the region of moderate
$Q^2$ relevant to neutrino experiments, where the strong
unequality of masses of the charmed and strange quarks
manifests its effects. Two important points are the onset
of the parton model interpretation of heavy flavors
when passing over the threshold value of $Q^2 \sim 4 \mu^2,
(m+\mu)^2$, and the transition from the
dominance of the longitudinal cross section
and the strong breaking
of the Callan--Gross relation at small $Q^2$ to the dominance of
the transverse cross section and the (approximate) restoration
of the Callan--Gross relation  at large $Q^2$.

To the lowest order in QCD, the longitudinal
and transverse absorption cross sections in the charged current
case are given by a generalization of the formulas
provided in \cite{NZ} -- see eqs.~(11) and (12) -- and
read ($\alpha_{ew}$ is the appropriate electroweak
coupling constant for the specific process considered)

\arr
\frac{d\sigma_{T,L}}{dz \, d^2 \mbox{\bm $k$}}
&=&
\frac{16}{\pi^2} \alpha_{ew} \,\alpha_{S}(\tilde{k}^{2})
\int d^{2}\mbox{\bm $\kappa$}\,
\frac{V(\mbox{\bm $\kappa$})\alpha_{S}(\mbox{\bm $\kappa$}^2)}
{(\mbox{\bm $\kappa$}^2+\mu_G^2)^2} \nonumber \\
&\times &
\left\{{N_{T,L}(\mbox{\bm $k$},\mbox{\bm $k$})
 \over[ \mbox{\bm $k$}^{2}+\varepsilon^{2}]^{2} } +
{N_{T,L}(\mbox{\bm $k$}+\mbox{\bm $\kappa$},
\mbox{\bm $k$}+\mbox{\bm $\kappa$})
 \over[(\mbox{\bm $k$}+\mbox{\bm $\kappa$})^{2}+\varepsilon^{2}]^{2}}
-2{N_{T,L}(\mbox{\bm $k$},\mbox{\bm $k$}+\mbox{\bm $\kappa$})
 \over [(\mbox{\bm $k$}+\mbox{\bm $\kappa$})^{2}+\varepsilon^{2}]
[\mbox{\bm $k$}^{2}+\varepsilon^{2}] }  \right\}
\label{eq:1}
\endarr
where the gluon-gluon-nucleon vertex function $V(\mbox{\bm $\kappa$})$
is related to the charge form factor of the proton
$G_{em}(\mbox{\bm $q$}^{2})$ through
$V(\mbox{\bm $\kappa$})= 1-G_{em}(3\mbox{\bm $\kappa$}^{2})$,
\beq
\varepsilon^{2}=z(1-z)Q^{2}+zm^{2}+(1-z)\mu^{2} \, ,
\label{eq:2}
\endeq
and the
functions $N_{T,L}(\mbox{\bm $k$}_{1},\mbox{\bm $k$}_{2})$
are explicitly given by
\arr
N_{T}(\mbox{\bm $k$}_{1},\mbox{\bm $k$}_{2}) &=&
[z^{2}+(1-z)^{2}](g_{V}^{2}+g_{A}^{2}) \mbox{\bm $k$}_{1}
\cdot \mbox{\bm $k$}_{2} \nonumber \\
& & +g_{V}^{2}[zm+(1-z)\mu]^{2}
+g_{A}^{2}[zm-(1-z)\mu]^{2}    \,\, ,
\label{eq:3}
\endarr

\arr
N_{L}(\mbox{\bm $k$}_1,\mbox{\bm $k$}_2) &=&
\frac{1}{Q^2}
\left(
[g_V^2(m-\mu)^2+g_A^2(m+\mu)^2]
\mbox{\bm $k$}_{1} \cdot \mbox{\bm $k$}_{2} \right. \nonumber \\
& &
+ g_V^2 \{ 2Q^{2} z(1-z)+(m-\mu)[zm-(1-z)\mu] \}^{2} \nonumber \\
& &
+ \left. g_{A}^{2} \{ 2Q^{2}z(1-z)+(m+\mu)[zm+(1-z)\mu] \}^{2}
\right) \, \, .
\label{eq:4}
\endarr
In muon scattering $g_{V}=e_{i}$ (the charge of the quark
involved in units of the electron charge), $g_{A}=0$ and $\mu=m$.
In the charged current (CC) neutrino interactions $g_{A}=-g_{V}=-1$
and $m$ and $\mu$ stand for the strange and the charm
quark masses. In
the neutral current (NC) neutrino interactions $\mu=m$ and the
corresponding vector and axial coupling are given by the Standard
Model \cite{LPR}. The strong coupling $\alpha_{S}(\tilde{k}^{2})$ in
front of the differential cross section enters at the virtuality of
the quark $\tilde{k}^{2}=\varepsilon^{2}+\mbox{\bm $k$}^{2}$.


The $z$--distributions of the charmed quark can be
calculated by integrating eq.~(\ref{eq:1}) over the
transverse momentum.

Let us start with the $z$-distribution for $\sigma_{T}$.
Here $N_{T}$ is a mild function of $z$, and
the $z$ dependence comes from $\varepsilon^{2}$. To a crude
approximation, neglecting the scaling violations, we find
\beq
{d\sigma_{T}\over dz}
\sim {z^{2}+(1-z)^{2} \over \varepsilon^{2}}=
{z^{2}+(1-z)^{2} \over z(1-z)Q^{2}+zm^{2}+(1-z)\mu^{2}}\, .
\label{eq:5}
\endeq
Consider first the muon scattering, $\mu=m$. The distribution
(\ref{eq:5}) is nearly flat at $Q^{2} \lsim 4\mu^{2}\sim 10
\,{\rm GeV}^{2}/c^2$. Only
at asymptotically high $Q^{2}\gg 4\mu^{2}$,  it develops the
parton model peaks
at $z\rightarrow 0$ and $z\rightarrow 1$ (see \cite{AM}), so that
the charmed (anti)quark carries a fraction $z\sim 1- \mu^{2}/Q^{2}$
of the photon's light-cone momentum. The result of the
exact calculation of $(1/\sigma_{T})
d\sigma_T/dz$ for the $\bar c c$ excitation in electromagnetic
scattering is shown in Fig.~2a.
The situation is similar for the NC neutrino scattering (Fig.~2b).

In the CC neutrino scattering the charmed quark is much heavier
than the strange quark, $\mu \gg m$.
At large $Q^{2}$
the $z$ distribution (shown in Fig.~2c)
again develops peaks at $z\rightarrow 1$ and
$z\rightarrow 0$. However,
the $Q^{2}$ evolution of the forward,
$d\sigma_{T}/dz \sim (1-z-m^{2}/Q^{2})$, and the backward,
$d\sigma_{T}/dz \sim (1-z-\mu^{2}/Q^{2})$,
peaks is quite different: the forward peak appears sooner.

Now, the CC excitation of charm and (anti)strangeness are
inseparable. In the parton model language, at
very large $Q^{2}$,
the peak at
$z\rightarrow 1$ can be identified with the excitation of the
charm on the strange sea, while the peak at $z\rightarrow 0$
describes the excitation of $\bar{s}$ on the $\bar{c}$ distribution
in the nucleon. Thus, only asymptotically the separation
of the two subprocesses depicted in Fig.~1 is well defined.
As we have already mentioned, in the
extraction of the dimuon data, a cut on $z$ is implicitly made,
which produces an acceptance--dependent separation of the
strange and charm contributions
to the $W$--absorption cross sections.
However, at moderate $Q^{2}$ the sensitivity to the cutoff $z_{c}$ is
rather strong and the parton model
reinterpretation of the $c\bar{s}$ excitation
in terms of the two partonic subprocesses is by no means
unique and becomes matter of convention.
In this region of $Q^2$ an assumption of the form
$d \sigma/dz \sim \delta (z-1)$,
in conjunction with a $Q^2$--independent fragmentation
function,  is untenable
because it would neglect the crucial $Q^2$--dependence
of the $z$--distributions.


The problem is complicated by the
presence of the longitudinal contribution,
which
contribute significantly to the ``non-partonic'' domain of
$z$ around $1/2$ (for a discussion of the massless
case see \cite{AM}).

In muon interactions, again neglecting the scaling violations,
one finds a broad symmetric $z$--distribution, shown in Fig.~2a
\beq
{d\sigma_{L} \over dz}
\sim {Q^{2}z^{2}(1-z)^{2} \over \varepsilon^{4}}\,\,,
\label{eq:7}
\endeq
which, in proximity of the forward peak, increases
with $Q^2$.
The ratio $R = \sigma_L/\sigma_T$ is of course rather small,
$R \sim Q^2/4 \mu^2$.

By contrast, in neutrino scattering,
because of the
nonconservation of the weak axial current ($g_{A}\neq 0$) and of
the flavor changing vector current ($\mu \neq m$), one has
$N_{L} \sim N_{T}\mu^{2}/Q^{2}$ and
the longitudinal contribution is rather significant at
moderate $Q^2$ ($R \sim 4 \mu^2/ Q^2$).
In the NC interactions the $z$-distribution is symmetric around
$z={1 \over 2}$ (Fig.~2b). In the CC interactions, $d\sigma_{L}/dz$
develops two peaks at $z\rightarrow 1$ and $z\rightarrow 0$, and is
asymmetric around $z={1\over 2}$ (Fig.~2c).



An interesting quantity is $R(z)=\sigma_{L}(z)/\sigma_{T}(z)$
which shows how the weak-current non-conservation effects depend
on $z$. In Fig.~3  we compare $R(z)$ at $Q^2= 4\, GeV^2/c^2$
for $\bar c c$ excitation in muon scattering and
in NC neutrino scattering, and for $\bar c s$ excitation in
CC neutrino scattering. Shown for comparison
are the values of $R$ for the $z$-integrated cross sections. In
virtual photoabsorption
both $R$ and $R(z)$ are small. In the NC and CC
neutrino scattering $R(z)$ decreases as $z,(1-z)
\rightarrow 0$.
Notice that in the CC case
$R(z)$ is slightly
asymmetric.

The peculiar $Q^2$--dependence of the transverse and
longitudinal contributions to the $z$--distributions
manifests itself in a detectable way at the level of
the sea parton densities.
We must say that
neither the introduction of the density of partons at
$Q^{2}\lsim 4\mu^{2},\,(m+\mu^{2})^{2}$, nor the comparison of the
$\sigma_{T}$-dominated muoproduction with the $\sigma_{L}$-dominated
neutrino--production do make much sense. Nonetheless, in order to
clarify the issue of the non--universality, let us proceed
with such a comparison.

The well-defined quantities
are the cross sections $\sigma_{T,L}$, which can be converted
into the structure functions $F_{L,T}(x,Q^{2})=Q^{2}\sigma_{L,T}/
4\pi\alpha_{ew}$. In terms of the more familiar
structure functions $F_1,F_2$, one has
$F_{1}=F_{T}/2x$ and
$F_{2}=F_{T}+F_{L}$.
At very large $Q^{2}$, when all quarks can be
regarded as massless, the structure functions can be decomposed in
terms of the parton densities as \cite{LPR} (we consider an isoscalar
nucleon)
\beq
F_{2}^{(\mu)}(x,Q^{2})=
{4\over 9}[u_{\mu}+\bar{u}_{\mu}+c_{\mu}+\bar{c}_{\mu}]
+{1\over 9}[d_{\mu}+\bar{d}_{\mu}+s_{\mu}+\bar{s}_{\mu}]
\label{eq:8}
\endeq
\beq
F_{2}^{(\nu)}(x,Q^{2})=
u_{\nu}+\bar{u}_{\nu}+ 2 \, \bar{c}_{\nu}+
d_{\nu}+\bar{d}_{\nu}+2\, s_{\nu}
\label{eq:9}
\endeq
These formulas can be taken as the operational definition of the parton
densities even at $Q^{2} \lsim 4\mu^{2},\,(m+\mu)^{2}$.
Similarly, one can introduce the parton densities $q_{\nu,\mu}^{(T)}$
defined in terms of $F_{T}(x,Q^{2})$, which would have been identical
to the ones in Eqs.~(\ref{eq:8},\ref{eq:9}) were it not for the
breaking of the Callan-Gross relation. Notice that
we have supplied the parton
densities by subscripts $\nu$ and $\mu$. The
non--universality of the charm and strange densities
is quantified by the ratio
$r_{\nu/\mu}=(\bar c_{\nu}+s_{\nu})/(\bar c_{\mu}+
s_{\mu})$ (however
one has to keep in mind that there is no direct
measurement of $s_{\mu}$).
We have evaluated $r_{\nu/\mu}$ within the model of
Ref.~\cite{BGNPZ1a,BGNPZ1b,BGNPZ1c,BGNPZ1d,BGNPZ2}
taking $(m+ \mu)^2 = 4 \, GeV^2/c^2$.
The result is shown in Fig.~4a, where
both the $x$-- and the $Q^2$--dependence of $r_{\nu/\mu}$
is exhibited. Notice that, whereas at large $Q^2$ the ratio
$r_{\nu/\mu}$ tends to the asymptotical (and naively assumed)
value of unity, at small $Q^2$ ($\lsim 10 \, GeV^2/c^2$)
it increases with $Q^2$ before flattening down.
In the kinematical region probed by the CCFR experiment
\cite{CCFR} the deviation of $r_{\nu/\mu}$ from unity
is as large as $\sim 25 \%$ and is a possible explanation
of the observed discrepancy between different determinations
of the strange density \cite{BGNPZ1b}.
The numerator in $r_{\nu/\mu}$ is dominated by the longitudinal
component: taking only the transverse contribution, the departure
of $r_{\nu/\mu}$ from unity would be even stronger.
For completeness
we present in Fig.~4b the ratio $c_{\mu}/(c_{\mu}+s_{\mu})$
evaluated at two different values of $Q^2$.

Whereas in the determination of $F_2^{(\nu)}$
one deals with the sum $\bar c_{\nu} + s_{\nu}$, which is
free from ambiguities, being related to the integral
of $d \sigma_{T,L}/dz$ over the whole $z$--range, the experimental
analysis of the dimuon data introduces a separation of
$s_{\nu}$ and $\bar c_{\nu}$. The $c \bar s$ production
cross section is thus splitted between the two partonic subprocesses
of Fig.~1.
This amounts to introducing a
cutoff $z_c$ in the $z$--distributions, so that
$s_{\nu}$ and $\bar{c}_{\nu}$
are defined
in terms of the $c\bar{s}$ production
cross section subject to the (arbitrary) cuts
$z>z_{c}$ and $z<z_{c}$, respectively.
In the treatment of the dimuon data
the cutoff $z_c$ is implicitly posited and the
strange density consequently obtained.
Were the data taken at asymptotically large $Q^2$, when
the peak at $z=1$ is delta--like, the choice of the cutoff
would be irrelevant. However, the $Q^2$ region of interest,
especially at small $x$,  is that
of small and moderate values $Q^2 \lsim 20 \, GeV^2/c^2$, where
the choice of $z_c$ necessarily introduces some arbitrariness.

Choosing the cutoff $z_c= 1/2$ is equivalent, in the muon
interaction and in the weak NC interaction, to simply equating the
charm and anticharm densities, since the $z$--distributions
are symmetric. Also in the CC interaction at asymptotically
large $Q^2$, where the two parton model peaks tend to become
delta--like, taking $z_c=1/2$ gives $s_{\nu}= \bar c_{\nu} =
(\bar c_{\nu} + s_{\nu})/2$. By contrast, in the CC case
at small $Q^2$ the $z$--distributions are strongly asymmetric , the
backward and the forward peaks are barely visible and the
splitting of $s_{\nu}$ and $\bar c_{\nu}$ cannot be unambiguously
done. Hence some care must be exerted in analyzing the dimuon
data in terms of the strange density. The effect of two
different choices of the cutoff $z_c$ on $s_{\nu}$
($z_c=1/2$ and $z_c=0.8$, the latter corresponding approximately
to taking $s_{\nu}= \bar c_{\nu} = (s_{\nu} + \bar c_{\nu})/2$)
is displayed in Fig.~5. Our curves are presented for
$Q^2 = 10\, GeV^2/c^2$ and compared to the dimuon data
around the same value of $Q^2$.
An exact comparison with the data would require
the knowledge of the experimental value of $z_c$, which is
lacking.

In conclusion, the calculation of the energy--momentum distribution
of the quarks produced in the charm leptoproduction
has shown that, in the presently explored kinematical
region, mass and current non--conservation effects
play a decisive role, making the extraction of the
heavy--quark parton densities from dimuon data subject to some
(solvable) ambiguity.
We have expressed in a quantitative way the deviations from the
naive parton model expectations and stressed the importance of a correct
interpretation of the data coming from neutrino
interactions.

\pagebreak

\pagebreak

\begin{center}

{\bf\large  Figure captions}

\end{center}

\begin{itemize}

\item[Fig.~1 - ]
The excitation of charm in the $W$--gluon fusion
process.


\item[Fig.~2 - ]
{\it a)} The $z$--distributions $(1/\sigma_{T,L}) d \sigma_{T,L}/
d z$ of the charmed quark produced in
the transverse (T) and the longitudinal (L) virtual
photoabsorption. The curves are as follows:
T at $1 \, GeV^2/c^2$, dotdashed; T at
$10 \, GeV^2/c^2$, solid; L at $1 \, GeV^2/c^2$, dotted; L at
$10 \, GeV^2/c^2$, dashed.\\
{\it b)} Same as {\it a)},
but for the weak neutral current case ($Z^0$--absorption
cross sections).\\
{\it c)} Same as {\it a)},
but for the charged current case ($W$--absorption
cross sections).

\item[Fig.~3 -] The ratio $R(z) = \sigma_L(z)/\sigma_T(z)$
at $Q^2 =4 \, GeV^2/c^2$ in muon scattering (solid curve),
CC neutrino scattering (dashed curve), NC neutrino scattering
(dotdashed curve).

\item[Fig.~4 - ]
{\it a)} The ratio $(\bar c_{\nu} + s_{\nu}) / (\bar c_{\mu} + s_{\mu})$
 plotted as a function of $x$ at three
different values of $Q^2$ (solid curve, $4 \, GeV^2/c^2$;
dashed curve, $10 \, GeV^2/c^2$; dotdashed curve, $30 \,
GeV^2/c^2$).\\
{\it b)} The ratio $c_{\mu}/(c_{\mu}+s_{\mu})$ at two
different valus of $Q^2$: $10 \, GeV^2/c^2$ (dashed curve),
$30 \, GeV^2/c^2$ (dotdashed curve).

\item[Fig.~5 - ]
The strange density $s_{\nu}$ probed in neutrino interactions
calculated at $Q^2 = 10 \, GeV^2/c^2$
with two different cutoffs: $z_c=0.5$, solid curve (T+L),
dashed curve (only T); $z_c=0.8$, dotdashed curve (T+L), dotted
curve (only T). The data are from Ref.~[5].

\end{itemize}

\end{document}